\documentstyle[emulateapj]{article}
\lefthead{Alonso-Herrero et al.}
\righthead{The [Fe\,{\sc ii}] emission in M82 and NGC~253}

\begin{document}
\title{The [Fe\,{\sc ii}]$1.644\,\mu$m  emission in
M82 and NGC~253: Is it a measure of the 
supernova rate?\footnote{Based on observations with the NASA/ESA Hubble
Space Telescope, obtained at the Space Telescope Science Institute, which
is operated by the Association of Universities for Research in Astronomy,
Inc. under NASA contract No. NAS5-26555.}}

\author{Almudena Alonso-Herrero, George H. Rieke, Marcia J. Rieke
and Douglas M. Kelly} 
\affil{Steward Observatory, The University of Arizona, Tucson, AZ 85721}

\begin{abstract}
We present {\it HST}/NICMOS [Fe\,{\sc ii}]$1.644\,\mu$m, Pa$\alpha$
($1.87\,\mu$m) and continuum images of the starburst galaxies M82 and
NGC~253 at an unprecedented
spatial resolution. In both galaxies  
we detect [Fe\,{\sc ii}] compact sources superimposed on
a diffuse background in the disk of the galaxies together with a
component above and below the plane of the galaxy. 
The radio and [Fe\,{\sc ii}] emissions
perpendicular to the disk of M82 show a remarkable similarity to
each other, suggesting 
that both emissions originate in shocks from supernova
explosions. We find a spatial  
correspondence between bright {\it compact}
[Fe\,{\sc ii}] emitting regions 
and the location of radio supernova remnants (SNR) for approximately 
$30-50$\% of radio SNRs 
in  M82 and NGC~253. This lack of a one-to-one correspondence, 
more than being indicative of a different origin for the radio and 
[Fe\,{\sc ii}] emission in starbursts,  suggests 
two populations of SNRs: an older population
($ \lesssim 10^4\,$yr) traced by the [Fe\,{\sc ii}] emission
and a younger population (a few hundred years old) traced by the radio SNRs.
We therefore conclude that the [Fe\,{\sc ii}] emission in starburst 
galaxies provides a good estimate of the supernova activity. 
Using our newly determined 
[Fe\,{\sc ii}] luminosities (corrected for extinction) of M82 and NGC~253,
we reevaluate the calibration of the supernova rate in
terms of the [Fe\,{\sc ii}] luminosity for starburst galaxies.

\end{abstract}


\section{INTRODUCTION}
Compact radio sources in nearby starburst galaxies often
appear to be supernova remnants (SNR). SNR in our 
Galaxy and the Large Magellanic 
Cloud (LMC)  are sources of bright near-infrared [Fe\,{\sc ii}] line emission
with luminosities ranging from $10^{26}\,$W to $3\times10^{29}\,$W
(for the [Fe\,{\sc ii}]$1.644\,\mu$m line, e.g., 
Oliva, Moorwood, \& Danziger 1989; Burton \& Spyromilio 1993).
This behavior suggests a connection between supernova
activity and the [Fe\,{\sc ii}] emission in starbursts. This is supported by 
near-infrared spectroscopy (Moorwood \& Oliva 1988;
Greenhouse et al. 1991; Vanzi \& Rieke 1997), by high-resolution
[Fe\,{\sc ii}] imaging studies (Forbes et al.
1993; Greenhouse et al. 1997; Alonso-Herrero et al.
2000), and by the tight correlation
between the radio 6\,cm and the [Fe\,{\sc ii}] emissions (Forbes \& Ward 1993).
Further determination of the relation of the infrared [Fe\,{\sc ii}] emission
with the supernova activity in nearby
starforming galaxies would be helpful both for the understanding
of these galaxies and for using the near-infrared [Fe\,{\sc ii}] luminosity
as estimator of the supernova rate in more distant galaxies
where individual radio supernovae cannot be identified.
                                            
There are indications of some complexities in this relation.
For example, the bright [Fe\,{\sc ii}] emission
may not be spatially coincident with the radio SNR.
Lumsden \& Puxley (1995) conducted a spectroscopic
survey for [Fe\,{\sc ii}] emission from radio SNRs in M33 and
found that in some cases the [Fe\,{\sc ii}] emission
does not peak at the position of the
radio SNR, as well as that some radio SNR show no
[Fe\,{\sc ii}] emission. Recently, Morel, Doyon, \& 
St-Louis (2002) have carried out an [Fe\,{\sc ii}] imaging survey of 
optically selected SNR in M33 and found a similar result.  
Both works have concluded that the
[Fe\,{\sc ii}] emission from SNR may be produced over a
time scale of $\sim$ 10$^4$ years.
Greenhouse et al. (1997) found that none of the brightest 
sources of [Fe\,{\sc ii}]
emission in M82 were coincident with the position of the radio SNRs.

M82 and NGC~253 are the nearest starburst galaxies (distances $d = 3.2\,$Mpc
and $d = 2.5\,$Mpc respectively). Both
galaxies show a large number of radio compact
sources (e.g., $\simeq 50$ in M82 Kronberg et al. 1985 and
Huang et al. 1994; $\simeq 60$ in NGC~253 Turner \&
Ho 1985 and Ulvestad
\& Antonucci 1991, 1997 and references therein). Their proximity
makes these two galaxies
ideal to study the relation between supernova activity and
the [Fe\,{\sc ii}] emission in starburst galaxies.
Greenhouse et al. (1997) presented a Fabry-Perot
[Fe\,{\sc ii}]$1.644\,\mu$m map  for M82 with a spatial resolution
of $\simeq 1.3 \arcsec$. Their main conclusions were that the compact
[Fe\,{\sc ii}] sources contribute  only $\simeq 14\%$ of the
total [Fe\,{\sc ii}] emission in M82 and that they trace a population of SNR
older than the radio SNR. Their work is somewhat inconclusive because
of the  limited angular resolution and also because strong extinction
could hide [Fe\,{\sc ii}] emission from some of the radio SNRs.
This paper presents {\it HST}/NICMOS images of M82 and NGC~253. 
The high spatial resolution 
of the NICMOS  [Fe\,{\sc ii}]$1.644\,\mu$m images 
(${\rm FWMH} = 0.35\arcsec$; this corresponds
to  5\,pc and 4\,pc for M82 and NGC~253, respectively) allows a
detailed study of the location of the bright sources of
[Fe\,{\sc ii}]  and of the r\^ole of supernova remnants in this line 
emission.

\section{NICMOS OBSERVATIONS}

\subsection{Data Reduction}
{\it HST}/NICMOS observations were
obtained on April 7, 8, 12 and 22 1998 (for M82) and
on August 13 and 20 1998 (for NGC~253) using the NIC2 and NIC3 cameras.
Table~1 lists details of the observations.
The observational strategy consisted of taking images in
a spiral dither with a 5.5 pixel spacing with
four  positions for a given field. The plate scales for the 
NIC2 and NIC3 cameras 
are 0.076\arcsec\,pixel$^{-1}$ and 0.20\arcsec\,pixel$^{-1}$,  
respectively. The FWHM of the point sources in the NIC3 F166N continuum 
images is $0.35\arcsec$.

\begin{deluxetable}{lcclccc}
\tablewidth{13cm}
\footnotesize
\tablecaption{Log of the HST/NICMOS observations.}
\tablehead{\colhead{Camera} & \colhead{Filter} & \colhead{line/filter} &
\colhead{galaxy} & \colhead{$t_{\rm exp}$} & \colhead{no. fields}
& \colhead{field of view}\\
\colhead{(1)} & \colhead{(2)} & \colhead{(3)} &
\colhead{(4)} &\colhead{(5)} & \colhead{(6)} &\colhead{(7)}}
\startdata
NIC2   & F160W & $H$ & M82 & 192 & 4 & $29.2\arcsec \times 72.7\arcsec$ \\
       &       &     & NGC~253 & 192 & 2 & $26.3\arcsec \times 33.1\arcsec$\\
NIC2   & F222M & $K$ & M82 & 448 & 4 & $29.2\arcsec \times 72.7\arcsec$\\
       &       &     & NGC~253 & 512 & 2 & $26.3\arcsec \times 33.1\arcsec$ \\
NIC2   & F187N & Pa$\alpha$ & M82 & 384 &
4 & $29.2\arcsec \times 72.7\arcsec$\\
       &      &             & NGC~253 & 384 & 2 &
$26.3\arcsec \times 33.1\arcsec$\\
NIC2   & F190N & continuum & M82 & 384 & 4 & $29.2\arcsec \times 72.7\arcsec$\\
       &       &           & NGC~253 & 384 & 2 &
$26.3\arcsec \times 33.1\arcsec$\\
NIC3   & F164N & [Fe\,{\sc ii}]$1.644\,\mu$m  &
M82 & 4096 & 2 &$53.8\arcsec \times 66.2\arcsec$\\
      &        &      & NGC~253 & 2560 & 1 & $51.2\arcsec \times 51.2\arcsec$\\
NIC3   & F166N & continuum & M82
& 3840 & 2 &$53.8\arcsec \times 66.2\arcsec$\\
       &      &            & NGC~253 &
2560 & 1 & $51.2\arcsec \times 51.2\arcsec$\\
NIC3   & F215N & continuum & M82 & 2304 & 2 &$53.8\arcsec \times 66.2\arcsec$\\
     &        &         & NGC~253 & 2304 & 1 &
$51.2\arcsec \times 51.2\arcsec$\\
\enddata
\tablecomments{Column~(1): NICMOS camera. Column~(2): Filter. Column~(3):
Corresponding ground-based broad-band filter, emission line or adjacent
continuum. Column~(4): Galaxy. Column~(5): total integration
time per field. Column~(6): Number of fields observed to produce
the final mosaic image. Column~(7): Field of
view of the mosaic before rotation to the common orientation
of north up, east to the left.}
\end{deluxetable}

The reduction of the NICMOS images used routines from the package
NicRed (McLeod 1997). A master dark image was produced by
combining between 10 and 20 darks for
a given sample sequence after the subtraction of the first readout.
The darks used were part of other proposals observed close in time,
and the flatfield images were on-orbit data. The data reduction
was performed using the following steps:
subtraction of the first readout, dark
current subtraction on a readout basis, correction for linearity and cosmic
ray rejection (using fullfit), and flatfielding. The individual
dithered galaxy images
were registered to a common position using fractional pixel offsets and
cubic spline interpolation, and
combined to produce the final image of each field.

\subsection{Flux-calibration of the images}

Prior to the flux calibration of the images, the background needs to be
subtracted (only for wavelengths longwards of $2\,\mu$m).
Due to the large projected size of both galaxies, the field of view
of the NIC2 images is not large
enough to enable measurements of the background on empty corners of the
images, so we used background measurements taken during
the Servicing Mission Observatory Verification (SMOV) program for the
filter NIC2 F222M. The F215N images cover a larger
field of view, and an estimate of the thermal background was obtained
from the corners of the images.
The flux calibration of the broad-band, on-line and off-line images
was performed using the conversion factors
(from ADU/s to mJy) based on measurements of the standard star P330-E during
SMOV (Marcia Rieke 1999 private communication).

Ideally one
would like to measure the continuum on both sides of the emission line
to estimate the continuum at the emission
line. However, NICMOS only provides continuum bands to the red of the
emission line.  In cases of objects such as
M82 and NGC~253, where the extinction to the nuclear regions is very
high (even at infrared
wavelengths), a straight subtraction of the longer wavelength off-line
image may result in  an over correction of the continuum at the
wavelength of the emission line. To estimate the continuum
for the NIC3 F164N images, we fitted the continuum on a pixel-by-pixel
basis between
$1.66\,\mu$m and $2.15\,\mu$m by linear regression of the flux as a function
of the wavelength for the NIC3 F166N and NIC3 F215N
line-free images. The extrapolated continuum was then subtracted from the 
NIC3 F164N image to produce the final continuum-subtracted 
[Fe\,{\sc ii}]$1.644\,\mu$m line image.  This procedure
does not correct for extinction,
but it rather ensures the subtraction of the correct {\it obscured} continuum
at each position of the galaxy. For the NIC2 F187N
image (Pa$\alpha$), since we did not have other narrow-band images taken with
camera 2, we used the NIC2 F190N continuum image. Fortunately, the
continuum need not be subtracted as accurately for Pa$\alpha$ as
for the [Fe\,{\sc ii}] emission line because the Pa$\alpha$ emission
is much brighter.

The fully-reduced images of each individual
field were combined to produce the mosaics.
The images presented in this paper were taken during different
orbits. This resulted in slightly different
orientations (from ${\rm P.A.} = -100\arcdeg$ to $- 105\arcdeg$
for M82 and from ${\rm P.A.} = 57\arcdeg$ to $68\arcdeg$ for
NGC~253). Before combining the images into mosaics, they 
were rotated by linear interpolation to north up, east to the left. This
provided the same orientation for all the images, and more importantly 
allows comparisons with previously published data.

To construct extinction maps to correct the
[Fe\,{\sc ii}] emission, and  [Fe\,{\sc ii}]/Pa$\alpha$ maps, the NIC2 F187N,
F190N, F160W and F222M images were demagnified to match the
plate scale of the NIC3 images and filtered with a Gaussian with
$\sigma = 0.65\,{\rm pixel} = 0.13\arcsec$ to match the
point spread function (PSF) of the NIC3 images.

\section{THE [Fe\,{\sc ii}]$1.644\,\mu$m AND Pa$\alpha$ MORPHOLOGIES}

\subsection{M82}
Figure~1a and 1b are mosaics (after rotation) of the
continuum at $1.66\,\mu$m and the continuum-subtracted
[Fe\,{\sc ii}]$1.644\,\mu$m line emission, respectively. Both images are
displayed  on a logarithmic scale to stress the contrast
between the diffuse extended continuum and the compact
sources. The [Fe\,{\sc ii}]$1.644\,\mu$m line
map  has been smoothed with a
Gaussian filter with $\sigma = 0.1\,\arcsec$.
The line emission image shows some
residuals which are due to both the under-sampling of the
NIC3 camera PSF and the variation in the shape
of the PSF with wavelength.
In addition, the right hand side of the line image shows a slightly lower
S/N because the F164N and F166N images
at that position had different orientations. 

The  [Fe\,{\sc ii}]$1.644\,\mu$m
line emission exhibits two distinct components: a number of relatively
compact regions distributed along the disk of the galaxy
and an extended diffuse component in the form of
filaments extending above and below the plane of the galaxy.
The filaments seen in the [Fe\,{\sc ii}]$1.644\,\mu$m emission line
show a remarkable similarity with the chimneys at the base
of the superwind detected in radio 
(Wills et al. 1999), which are interpreted by these authors as 
expelled material from the central region of M82.

In Figures~2a and 2b  we present contour maps of the
$1.66\,\mu$m continuum emission and the [Fe\,{\sc ii}]$1.644\,\mu$m line
emission, respectively. Comparison of the NICMOS
[Fe\,{\sc ii}]$1.644\,\mu$m line map with that of Greenhouse et
al. (1997) reveals that some of the
sources seen in the lower resolution map of Greenhouse et al.
(1997) break up into individual sources
at higher spatial resolution. In particular, their source
Fe1 contains some 4 or 5 individual sources. Source Fe6 (the most south-west
source in Greenhouse et al. 1997 map) appears as a point
source at our resolution in both the NIC3 F164N and NIC3 F166N images,
and the signal to noise on it is compromised by the differing image
orientations (see above). Therefore it is not surprising
we do not detect line emission from it.
The NICMOS images detect a number of
compact sources in addition to those reported in Greenhouse et al. (1997).

In Figures~2a and 2b we display the positions of 44 radio SNRs 
with peak  flux densities above 1\,mJy (list provided by Dr. Karen Wills, 
see more details in Section~5.1).
The distribution of the radio SNRs and their relation with the
[Fe\,{\sc ii}]$1.644\,\mu$m emission will be discussed in Section~5.

In Figures~3a and 3b we display the mosaics of the
$1.90\,\mu$m continuum  (NIC2 F190N images) and
the Pa$\alpha$ line emission (continuum subtracted NIC2 F187N
images) respectively.
The Pa$\alpha$ image is dominated by a ring of ionized gas with an
approximate diameter of 400\,pc. The resemblance with the
mid-infrared [Ne\,{\sc ii}] line emission map at $12.8\,\mu$m 
(Achtermann \& Lacy 1995) is remarkable.
The south side of this ring of ionized material appears to be incomplete
(see Achtermann \& Lacy 1995), and it also shows a ``hole'' of emission at the
center of the galaxy, which
approximately coincides with the peak of the
$2.2\,\mu$m surface brightness. 

\begin{deluxetable}{ccccccc}
\footnotesize
\tablecaption{Extinction to the nuclear regions of M82 using
a foreground dust screen model.}

\tablehead{
\colhead{Ratio: flux(1)/flux(2)} & \colhead{Ap.} &
\colhead{flux(1)} & \colhead{flux(2)} & \colhead{$A_V$} & \colhead{References flux(1), flux(2)}
%& (\arcsec) & (W m$^{-2}$) & (W m$^{-2}$) & (mag) &
}
\startdata
\multicolumn{6}{c}{Emission line measurements}\\
             &  (\arcsec) & (W m$^{-2}$) & (W m$^{-2}$) & (mag) \\ 
\hline
Pa$\alpha$/H$\alpha$ & 5.8 & $5.30 \times 10^{-15}$ &
$2.80 \times 10^{-15}$ & $5\pm0.5$ & this work, O'Connell \& Mangano (1978)\\
Pa$\alpha$/Pa$\beta$ & 6 & $5.30 \times 10^{-15}$ &
$5.90 \times 10^{-16}$ & $11\pm1$ & this work, McLeod et al. (1993)\\
Pa$\alpha$/Pa$\gamma$ & 3  & $1.27 \times 10^{-15}$ &
$0.5 \times 10^{-16}$ &
$10\pm1$& this work, McLeod et al. (1993)\\
Pa$\alpha$/Pa$\beta$ & 3 & $1.27 \times 10^{-15}$ &
$1.47 \times 10^{-16}$ &
$11\pm1$ & this work, McLeod et al. (1993) \\
Br$\gamma$/Pa$\alpha$  & 3.8 & $2.00 \times 10^{-16}$ &
$2.08\times 10^{-15}$ & $11\pm4$ & Lester et al. (1991), this work\\
Br$\alpha$/Pa$\alpha$ (E1) & 6? & $4.55 \times 10^{-15}$ &
$4.65\times 10^{-15}$ & $18\pm3$ & Achtermann \& Lacy (1995), this work\\
Br$\alpha$/Pa$\alpha$ (W1) & 6? & $3.92 \times 10^{-15}$ &
$5.92\times 10^{-15}$ & $13\pm2$ & Achtermann \& Lacy (1995), this work\\
Br$\alpha$/Pa$\alpha$ (W2) & 6? & $7.30 \times 10^{-15}$ &
$5.12\times 10^{-15}$ & $23\pm3$ & Achtermann \& Lacy (1995), this work\\
$[$Fe\,{\sc ii}]1.64/[Fe\,{\sc ii}]1.25 & 3 &
$9.20 \times 10^{-17}$ & $5.20 \times 10^{-17}$ &
$10\pm1$ & this work, McLeod et al. (1993)\\
\hline
\multicolumn{6}{c}{Continuum measurements}\\
             &  (\arcsec) & (mJy) & (mJy) & (mag) \\ 
\hline
F222M/F160W & 3 & 237 & 166 & $13\pm2$ & this work \\
\enddata
\tablecomments{Apertures are circular and 
centered at the peak of the $2.2\,\mu$m
continuum brightness, except for sources E1, W1 and
W2 in Achtermann \& Lacy (1995) notation. The approximate locations of 
the Br$\alpha$ (or Pa$\alpha$) sources from the peak of the near-infrared 
emission are: E1 (+4.1\arcsec, +0.5\arcsec), W1 ($-4.6$\arcsec, 
$-2.0$\arcsec) and W2 ($-10.3$\arcsec, $-4.5$\arcsec).
The errors quoted for the extinctions
include uncertainties from the aperture size and centering,
and flux calibration.}
\end{deluxetable}

\begin{deluxetable}{ccccccc}
\footnotesize
\tablecaption{Extinction to the nuclear regions of NGC~253 using
a foreground dust screen model.}
\tablehead{
\colhead{Ratio: flux(1)/flux(2)} & \colhead{Ap.} &
\colhead{flux(1)} & \colhead{flux(2)} & \colhead{$A_V$} & \colhead{References flux(1), flux(2)}
%& (\arcsec) & (W m$^{-2}$) & (W m$^{-2}$) & (mag) &
}
\startdata
\multicolumn{6}{c}{Emission line measurements}\\
             &  (\arcsec) & (W m$^{-2}$) & (W m$^{-2}$) & (mag) \\ 
\hline
Pa$\alpha$/Br$\gamma$ (center) & $3\arcsec \times 3\arcsec$ &
$1.58 \times 10^{-15}$ &
$1.41\times 10^{-16}$ & $5\pm7$ & this work, Harrison et al. (1998)\\
Pa$\alpha$/Br$\gamma$ (center) & $3.9\arcsec \times 4.8\arcsec$
& $2.40 \times 10^{-15}$ &
$2.35\times 10^{-16}$ & $11\pm6$ & this work, Puxley \& Brand (1995)\\
Pa$\alpha$/Br$\gamma$ (3\arcsec \ NE) & $3\arcsec \times 3\arcsec$ &
$1.20 \times 10^{-15}$ &
$1.15 \times 10^{-16}$ & $8\pm12$ & this work, Harrison et al. (1998)\\
Pa$\alpha$/Br$\gamma$ (3.9\arcsec \ NE) & $3.9\arcsec \times 4.8\arcsec$
& $2.00 \times 10^{-15}$ &
$1.83 \times 10^{-16}$ & $7\pm12$ & this work, Puxley \& Brand (1995)\\
\hline
\multicolumn{6}{c}{Continuum measurements}\\
             &  (\arcsec) & (mJy) & (mJy) & (mag) \\ 
\hline
F222M/F160W (center) & $3.9\arcsec \times 4.8\arcsec$
& 213 & 142 & $13\pm2$ & this work \\
\enddata
\tablecomments{Apertures are rectangular, and centered at 
the peak of the $2.2\,\mu$m
continuum brightness (center) and at 
the Pa$\alpha$ source located NE of the 
continuum peak. The errors in the derived extinctions take
into account 10\% and 20\% uncertainties for the nucleus and NE source
measurements, respectively.}
\end{deluxetable}

It has been suggested that higher extinction
in the south half of M82 may be causing the 
uneven distribution of  ionized gas in the ring. However, the ring appears
to be complete in
[Fe\,{\sc ii}]$1.644\,\mu$m line emission. If extinction were the
cause, we should expect to see a similar morphology in the [Fe\,{\sc ii}]
emission, which is more affected by extinction than the
Pa$\alpha$ emission. It is therefore tempting to suggest that the south part
of the ring of ionized gas
is actually {\it broken}, perhaps due to SN explosions.
This hypothesis would also explain the excess of [Fe\,{\sc ii}] emission in
this region. The presence of the hole of gas emission
at the center of M82 along with the
anti-correlation between the morphology of the hydrogen recombination
lines and the CO index has been invoked as evidence for
an outwards propagating starburst (e.g., Satyapal et al.
1997). The diffuse component of Pa$\alpha$ emission is also
in the form of ``fingers'' or ``filaments'' above and below the plane of
the galaxy. It is now known from ground-based H$\alpha$ imaging that
this emission along the minor axis of the galaxy
extends over 11\,kpc (Devine \& Bally 1999). Figure~1b and Figure~3b  show
how there is a very close spatial correspondence between the
morphologies of the extended
[Fe\,{\sc ii}]$1.644\,\mu$m and Pa$\alpha$ emissions.

\subsection{NGC~253}
Figures~4a and 4b are  the $1.66\,\mu$m continuum and
the [Fe\,{\sc ii}]$1.644\,\mu$m line emission maps of NGC~253. As
seen in M82, the [Fe\,{\sc ii}]$1.644\,\mu$m emission is in the form of a
number of compact regions along the disk of the galaxy superimposed
on a diffuse emission component. Figures~5a and 5b are contour plots
of these two emissions. We also show the positions of the 6\,cm radio sources
with $S_{\rm peak} > 1\,$mJy from Ulvestad \& Antonucci (1991), excluding
those sources that are classified as H\,{\sc ii} regions in 
Ulvestad \& Antonucci (1997).

Ground-based [Fe\,{\sc ii}]$1.644\,\mu$m and Br$\gamma$ maps at lower 
spatial resolution
(${\rm FWMH} \simeq 1.3\arcsec$) were previously observed by
Forbes et al. (1993). They found that the central [Fe\,{\sc ii}] emission
of NGC~253 is dominated by three peaks of emission. Our higher resolution
NICMOS image shows a larger number of [Fe\,{\sc ii}] compact
sources. For instance, source B (in the Forbes et al. 1993 notation)
breaks up into three individual  sources. In addition to the nucleus and
sources A and B detected by 
Forbes et al. (1993), there are other sources along the disk of the galaxy,
as well as diffuse emission perpendicular to the disk, more prominent
in the south half of the galaxy. We believe that Forbes et al.
(1993) did not detect the diffuse component above and
below the disk of the galaxy because of the lower S/N of their
[Fe\,{\sc ii}] map. We also mark in these figures the 
position of the peak (the filled square) 
of the near-infrared emission (also coincident
with the peak of the Pa$\alpha$ emission) which was identified 
as the nucleus of the galaxy by
Forbes et al. (1993). Other authors (e.g., Keto et al. 1999) claim that 
a bright non-thermal radio source located NE of the 
peak of the infrared continuum emission (source 2 in Turner \& Ho 1985) 
may be the true nucleus of 
NGC~253. It is shown in these figures as a filled star symbol.

The NICMOS Pa$\alpha$ line map (Figure~6b)
shows very bright emission stemming from the peak of infrared
continuum emission, together with more diffuse  emission in a
plume-like morphology along the plane of the galaxy, primarily to the
NE of the nucleus. As for M82, there is an extended component of the
Pa$\alpha$ emission perpendicular to the disk, better seen in our
NIC3 images (not shown here) which are more sensitive to diffuse emission. This
component is however not as strong as that in M82. There is a
remarkable resemblance of the central  Pa$\alpha$ emission with both the 
[Ne\,{\sc ii}]$12.8\,\mu$m emission line and the nearby 
continuum  images presented in 
B\"oker, Krabbe, \& Storey (1998) and Keto et al. (1999).

\section{EXTINCTION}

Both M82 and NGC~253 are  known to be highly obscured
(see for instance the comparison
between the optical {\it HST}/WFPC2 F555W and the
{\it HST}/NICMOS F222M images of
M82 in Alonso-Herrero et al. 2001). To calibrate
the supernova rate in terms of the
[Fe\,{\sc ii}]$1.644\,\mu$m emission
and to obtain the total number of ionizing photons, it is necessary to
correct the observed quantities for extinction.

\subsection{M82}
We can estimate the extinction to
the gas in M82 using observations of the ratios
of hydrogen recombination and near-infrared [Fe\,{\sc ii}] lines.
We used published values of the line fluxes for the
smallest apertures and matched those apertures on the NICMOS line images when
measuring the Pa$\alpha$ and [Fe\,{\sc ii}]$1.644\,\mu$m fluxes.
The intrinsic ratios of hydrogen recombination lines are from
Hummer \& Storey (1987) for an electron temperature of 
$T_{\rm e} =10^4\,$K and electron density of 
$n_{\rm e} =10^4\,$cm$^{-3}$. The ratio between
[Fe\,{\sc ii}]$1.257\,\mu$m and [Fe\,{\sc ii}]$1.644\,\mu$m
does not depend upon the physical conditions of the gas.
We used a polynomial fit to the
near-infrared extinction law given in He et al. (1995).

Since we do not have NICMOS images of two hydrogen recombination
lines, to measure extinction at the full NICMOS angular resolution
we made use of the broad-band filters NIC2 F160W
and NIC2 F222M to construct a map of extinction to the stars in M82.
We assumed a color for the unreddened stellar
population of $H-K = 0.2$ (which applies 
almost independently of the age of the population), and,
for computational convenience, a foreground dust screen model following
the extinction law of He et al. (1995). The contribution of dust emission at
the $K$-band was assumed to be negligible.
Figure~7a is the extinction map ($A_K$) from the $H-K$ color map of
M82, with the positions of the radio sources superimposed. The
values of the extinction in the $K$-band derived from this map are between
$A_K = 0.3\,$mag and $A_K = 2.5\,$mag. The M82
extinction map qualitatively looks very similar to those of Satyapal et
al. (1995), but quantitatively the values of the extinction are different.

The estimates of the extinction to the gas for M82 
are presented in the first part of Table~2 along with the line fluxes.
The errors in Table~2  are the uncertainties associated
with the aperture size, aperture centering and flux
calibration. For comparison, we also
estimated the extinction through a small aperture (3\arcsec)
using the F222M/F160W continuum flux ratio, which gives an
estimate of the visual extinction to the stars. As can be seen from this
table the agreement between the extinctions to the gas and to the stars
is excellent at infrared wavelengths. 
Satyapal et al. (1995) reached a similar conclusion. 
A much lower level of extinction is indicated by comparing
the H$\alpha$ and Pa$\alpha$ line fluxes. This behavior is
known to arise from the different optical depths probed by optical and
near-infrared lines. This behavior is confirmed by the much higher extinction
of $A_V \simeq 27\,$mag inferred by Puxley et al.
(1989) from the radio H53$\alpha$ to mid-infrared
Br$\alpha$ line ratio.
            
We also computed the extinctions to the E1, W1 and W2 regions
in Achtermann \& Lacy (1995) notation (see Table~2 for the positions). 
These were found to be bright in [Ne\,{\sc ii}]$12.8\,\mu$m and 
Br$\alpha$. We used  their Br$\alpha$
measurements along with the mid-infrared extinction curve in
Rieke \& Lebofsky (1985). Since Achtermann \& Lacy (1995) did not quote
aperture sizes, we assumed that their fluxes were measured over
the extent of these regions (approximately $5\arcsec-6\arcsec$ \ in diameter).
The values of the extinction for these three regions are slightly larger than the values
we would measure from the $H-K$ extinction map, perhaps indicating minor
differences in the optical depth. 

A further test of the extinction to the gas was based on the
total number of ionizing photons from the extinction
corrected Pa$\alpha$ line emission map. The number of ionizing
photons obtained from the extinction-free H53$\alpha$
radio recombination line is $N_{\rm Ly} = 1 \times 10^{54}\,$s$^{-1}$
(Puxley et al. 1989), and from the $3.3\,$mm continuum is
$N_{\rm Ly} = 1.1 \times 10^{54}\,$s$^{-1}$
(Carlstrom \& Kronberg 1991). We measured through a large aperture
the Pa$\alpha$ flux on the observed  and the
extinction-corrected images. We obtained
$N_{\rm Ly} = 4 \times 10^{53}\,$s$^{-1}$ (observed) and
$N_{\rm Ly} = 1.4 \times 10^{54}\,$s$^{-1}$ (corrected for extinction),
for electron temperature $T_{\rm e} = 10^4\,$K. The total number of
ionizing photons corrected for extinction is in excellent agreement with
a number of estimates including those of  Puxley et al. (1989), 
Carlstrom \& Kronberg (1991) and F\"orster Schreiber et al. (2001).

\subsection{NGC~253}                         
As for M82, the extinction to the stars was computed from the 
$H-K$ color map of NGC~253 (Figure~7b). It shows 
values between $A_V = 3\,$mag and $A_V=21\,$mag and
is very similar to that of
Sams et al. (1994) derived from a $J-K$ color map. We find a similar
value of the extinction ($A_V \simeq 21\,$mag) for the point with the
highest value of the extinction which was identified by Sams et al.
(1994) as the true nucleus of NGC~253.

To calculate the extinction to the gas in NGC~253 we used Pa$\alpha$ and
Br$\gamma$ fluxes (the latter from Puxley \& Brand
1995 and Harrison et al. 1998). Both works centered their
apertures on the approximate
coordinates of the secondary near-infrared peak and therefore we will assume
that the largest Br$\gamma$ fluxes in these works correspond to the
primary near-infrared peak. If their (0,0) position
were to be identified with the peak of the continuum emission, 
their fluxes would not be consistent with Forbes
et al. (1993)  Br$\gamma$ line emission map.
This map and our Pa$\alpha$ line emission map (Figure~6b) show that 
the brightest hydrogen line 
emission arises from the peak of the continuum emission.  
In Table~3 we give estimates for the extinction for the
primary near-infrared peak (center) and secondary near-infrared
peak (positions 3\arcsec \ and 3.9\arcsec \ to the NE of the
primary peak for Puxley \& Brand 1995 and
Harrison et al. 1998 works respectively). The errors take
into account 10\% and 20\% combined
flux calibration/centering uncertainties for the nucleus 
and NE source, respectively. 

The values of the gas extinction
to these two regions as derived from
the Pa$\alpha$/Br$\gamma$ line ratio are consistent (to within the errors)
with the extinction to the stars from the $H-K$ color map. Despite this
apparent good agreement we caution the reader of the
aforementioned centering uncertainties. We also measured
the observed and corrected for extinction Pa$\alpha$ fluxes  which
yielded a number of ionizing photons of $N_{\rm Ly} = 2.5 \times
10^{52}\,{\rm s}^{-1}$ (observed) and $N_{\rm Ly} = 1.4 \times
10^{53}\,{\rm s}^{-1}$ (corrected for extinction), for the 
field of view shown in Figure~6b. The latter
value is consistent with Engelbracht et al. (1998) and
Keto et al. (1999) estimates of the number of ionzing photons from
their large aperture Br$\gamma$ and [Ne\,{\sc ii}] measurements, but
smaller than Puxley et al. (1997) estimate ($N_{\rm Ly} = 3.7 \times
10^{53}\,{\rm s}^{-1}$) from millimeter-wavelength hydrogen recombination
lines.

\section{THE  [Fe\,{\sc ii}]$1.644\,\mu$m EMISSION OF M82 and NGC~253}

\subsection{The Distribution of the [Fe\,{\sc ii}]$1.644\,\mu$m
and radio SNRs}

To determine whether there is a spatial correspondence between
the compact radio sources and the [Fe\,{\sc ii}] sources in both M82
and NGC~253, very precise coordinates are needed. Unfortunately,
due to problems with the {\it HST} telemetry during the observations of
M82, and 
the uncertainties when producing and rotating the 
mosaics, the coordinates of the NIC3 images may not be
completely reliable. In addition, it is expected that there
are minor discrepancies between the radio and the optical/infrared
coordinate reference systems.

Our approach was to determine the coordinates of the
[Fe\,{\sc ii}] image by trying to find the highest possible number
of coincidences between the positions of the radio SNRs
and the [Fe\,{\sc ii}] sources.
We use DAOFIND within the DAOPHOT package in {\sc iraf}
to find compact sources. 
The term ``compact'' is not used in the sense of point sources.
In fact, most of these sources seem to be slightly 
resolved, although due to the under-sampling of the NIC3
PSF, some uncertainties remain. The sources were selected
to be easily distinguishable from the extended diffuse emission detected
above and below the plane of the galaxy.

In M82 we identified 65 ``compact'' sources in the
[Fe\,{\sc ii}]$1.644\,\mu$m map (not corrected for extinction).
For the radio sources in M82 we used the positions of SNRs
identified by searching for compact features 
within radio maps at 408\,MHz, 1.4\,GHz, 5\,GHz and 
8.4\,GHz, with peak 
flux densities above 1\,mJy (Dr. Karen Wills, private communication
2000). We excluded from this list radio sources that are classified 
as H\,{\sc ii} regions based on their radio spectral index (see McDonald et 
al. 2002 for details), resulting in 44 radio SNRs.
We began the comparison of positions using 
the coordinates in the header of the NICMOS NIC3 F164N images and made
small offsets of the [Fe\,{\sc ii}] image, cross-correlating the
positions of the [Fe\,{\sc  ii}] sources with the radio SNRs.
Based on the resolution of the NIC3 images and the
average angular size of the radio SNRs ($\simeq 0.2\arcsec$, Huang et al.
1994), we counted a coincidence when the angular separation between
a radio source and an [Fe\,{\sc ii}] source was 
$\leq 0.5$\arcsec. With a small shift, we found 21 coincidences between
radio SNRs and [Fe\,{\sc ii}] sources, whereas if we used
the original coordinates we could only find 7 coincidences.
Monte Carlo simulations showed that the probability of finding 21 
coincidences by pure chance is $2 \times 10^{-6}$ 
(see Whitmore \& Zhang 2002 for details on the method), 
so we believe that now the infrared and the radio images are properly
aligned. 

For NGC~253 we found discrepancies between the coordinates
derived for the near-infrared peak of
emission from the NIC2 and NIC3 images (which were taken
during two different orbits). Although no problems
were reported for these two sets of observations, some error must
be affecting the coordinates given in the headers
of the images. Again as with M82, we cross-correlated the positions
of the [Fe\,{\sc ii}]$1.644\,\mu$m (36 sources) and 35 radio sources 
taken from Ulvestad \& Antonucci 1991 ---excluding
those sources that are classified as H\,{\sc ii} regions in 
Ulvestad \& Antonucci (1997)--- using the coordinates
of the NIC3 images. We find a total of 9 coincidences for a separation 
between the radio sources
and [Fe\,{\sc ii}] sources $\leq 0.5\arcsec$. The probability of finding
9 coincidences in by pure chance in NGC~253 
is approximately $9 \times 10 ^{-3}$.

One important conclusion, despite the uncertainties of the image
alignment, is that
we cannot assign an [Fe\,{\sc ii}] source to each radio SNR or
vice versa. We find approximately $30-50$\% of radio SNRs 
in  M82 and NGC~253 have an [Fe\,{\sc ii}]. 
This point is illustrated in the contour
images of the [Fe\,{\sc ii}]$1.644\,\mu$m
line emission and continuum emission at $1.66\,\mu$m in Figures~2a
and 2b for M82 and Figures~5a and 5b for NGC~253. In all 
figures we show  the positions of the radio SNRs as asterisks. 

We also tried to determine if there is a relation between the presence 
of a coincidence and the age of the radio SNR. Such a correlation 
would be expected if the [Fe\,{\sc ii}]
emission were brighter in older supernova remnants.
For M82  we used the 
angular sizes of the radio SNR as an indication of their age 
(see Huang et al. 1994). In general, we did not find that older radio
SNRs tend to have a coincidence with an [Fe\,{\sc ii}] source.

\subsection{Luminosity and Origin of the [Fe\,{\sc ii}]$1.644\,\mu$m sources}

\subsubsection{Supernova Excitation}

We performed aperture photometry on the [Fe\,{\sc ii}]
sources identified in M82\footnote{Since 
the photometry was done on the [Fe\,{\sc ii}] image
registered to match the field of view and 
position of the Pa$\alpha$ image, the 
number of detected sources (71) is slightly 
different from that in Section~5.1} and NGC~253
using a fixed $0.8\arcsec$ diameter aperture. This aperture 
corresponds to a physical size of 12\,pc and 10\,pc for M82 and 
NGC~253, respectively.
Figures~9a and 9b show the distributions  of the measured
[Fe\,{\sc ii}]$1.644\,\mu$m luminosities (not corrected for extinction).
Some of the sources may be larger
than the aperture used for the photometry; on the other hand, patchy
extinction will make some of the sources appear smaller.
From inspection of typical source sizes, however, the
luminosities would be accurate on average. Recent studies of 
Galactic SNR show that the [Fe\,{\sc ii}]
emission is filamentary and have sizes of the order 
of 11\,pc (e.g, IC443 Rho et al. 2001; see also Morel et al. 
2002 for SNR in M33).

For the assumed distances to M82 and NGC~253 the
observed [Fe\,{\sc ii}]$1.644\,\mu$m luminosities of these
sources range from  $2.2 \times 10^{27}\,$W to $1.2 \times 10^{30}\,$W.
As a consistency check for the photometry, we measured the
[Fe\,{\sc ii}]$1.644\,\mu$m fluxes for the Fe1, Fe2 and Fe3 sources in M82 
identified by Greenhouse et al. (1997) 
using the aperture diameters listed in their table~1.
The agreement with Greenhouse et al. (1997) fluxes for
these sources is excellent, to within $3\%$. However, when we
compared our photometry for the sources reported in Forbes et al.
(1993) for NGC~253 we found some discrepancies.

The corrected [Fe\,{\sc ii}]$1.644\,\mu$m luminosities for compact
sources in M82 and NGC~253 range  from
$3 \times 10^{29}\,$W to $2.2 \times 10^{31}\,$W. As a comparison,
Galactic SNRs yield  [Fe\,{\sc ii}]$1.644\,\mu$m luminosities of
$1 \times 10^{26}\,$W to $7 \times 10^{28}\,$W (Oliva et al. 1989;
Keller et al. 1995), SNRs in the LMC yield luminosities 
from  $1 \times 10^{28}\,$W to
$3 \times 10^{29}\,$W (Oliva et al. 1989) and  SNRs in M33
(assuming a distance of 720\,kpc)
$5 \times 10^{27}\,$ to $3 \times 10^{29}\,$W
(Lumsden \& Puxley 1995 and Morel et al. 2002).

The individual sources account for some
22\% and 27\% of the total [Fe\,{\sc ii}]$1.644\,\mu$m emission (extinction
corrected) measured in M82 and NGC~253, respectively.
Given the estimates that the [Fe\,{\sc ii}] emitting
phase of a SNR has a duration of $\simeq 10^4\,$yr (Lumsden \&
Puxley 1995; Morel et al. 2002) and the supernova
rates in M82 and  NGC~253 are $\simeq 0.10\,$yr$^{-1}$ (Huang et al. 1994
and Ulvestad \& Antonucci 1997), the detected [Fe\,{\sc ii}] 
sources account for $\simeq 1000\,$yr worth of supernovae. Presumably
much of the remaining unresolved [Fe\,{\sc ii}] arises from
additional supernovae that are packed so densely that they are confused
in our images. Also, SNRs in these two galaxies
may expand and merge into a general hot ISM within a few thousand
years losing their identity as individual sources. Photometry on the 
[Fe\,{\sc ii}] emission arising from the disk of M82 shows that 
as much as 70\% of the total emission (disk plus component perpendicular 
to the disk) is associated with SNRs. This 
together with the good agreement between the radio and [Fe\,{\sc ii}]
morphologies of the components perpendicular to the disk gives further
indication for a SNR origin of the [Fe\,{\sc ii}] emission in 
starburst galaxies. 

    Lumsden \& Puxley (1995) pointed out that one possibility is that the
[Fe\,{\sc ii}] lines reach maximum luminosity when the remnant as a whole
undergoes the transition from adiabatic expansion to radiative expansion.
According to Shull \& Draine (1986), this happens when radiative cooling
occurs faster than dynamical times, around $10^{4}\,$yr after
the supernova explosion. Therefore if
the brightest [Fe\,{\sc ii}] emission occurs at approximately 
 $10^{4}\,$yr, it is not
surprising that we do not find a correspondence for all the
[Fe\,{\sc ii}] sources and all the radio
SNRs. Muxlow et al. (1994) found that half of the
radio sources in M82 are unresolved indicative of a young age, but 
the other resolved half are probably intermediate age. This together with
the fact that we found that $30-50\%$ of the radio SNR have an 
[Fe\,{\sc ii}] counterpart implies that the [Fe\,{\sc ii}] sources may 
be $\lesssim 10^{4}\,$yr.                                     

\subsubsection{Excitation in H\,{\sc ii} Regions}

One possibility for the lack of
one-to-one coincidences between the radio SNRs and the [Fe\,{\sc ii}]
sources would be that most of the [Fe\,{\sc ii}] emission in these 
two galaxies is 
produced in H\,{\sc ii} regions (photoionization). To check this, 
we measured the Pa$\alpha$ fluxes for the sources detected in
the [Fe\,{\sc ii}] extinction corrected map of M82, and 
assumed an [Fe\,{\sc ii}]$1.644\,\mu$m/Pa$\alpha$
ratio for H\,{\sc ii} regions as that observed in Orion. We found that only
$\simeq 6-8\%$ of the total [Fe\,{\sc ii}]$1.644\,\mu$m arises from
H\,{\sc ii} regions. This further supports the hypothesis that
most of the [Fe\,{\sc ii}]$1.644\,\mu$m emission along the
plane of M82 is produced by SNRs.

The [Fe\,{\sc ii}]/Br$\gamma$ ratios in
H\,{\sc ii} regions and SNR differ by a factor of a hundred or more,
and therefore can be used to indicate the excitation
mechanism of the [Fe\,{\sc ii}] line within our images.
In Alonso-Herrero et al. (1997)
we showed that the behavior of the [Fe\,{\sc ii}]$1.644\,\mu$m/Br$\gamma$
line ratio in starbursts and active galaxies
can be understood as a progression from pure photoionization
to pure shock excitation going from pure H\,{\sc ii} regions through
starbursts, transition objects, Seyferts to supernova remnants.

Figures~8a and 8b are the observed 
[Fe\,{\sc ii}]$1.644\,\mu$m/Pa$\alpha$ maps for M82 and NGC~253, respectively.
The values of the  [Fe\,{\sc ii}]$1.644\,\mu$m/Pa$\alpha$ ratio range from
0.02 in black (mostly along the disk of the galaxy) to 0.8 in white 
(some of
the bright [Fe\,{\sc ii}] sources). These ratios translate into
[Fe\,{\sc ii}]$1.644\,\mu$m/Br$\gamma$ line ratios of between 0.24 and 
10 (see also the histograms in Figure~10).
Since these maps are not corrected for extinction, these numbers
are actually lower limits. When the positions of radio sources
are plotted  in the [Fe\,{\sc ii}]$1.644\,\mu$m/Pa$\alpha$ maps
we find again that there is a number of coincidences
between the radio SNRs and the regions of
enhanced [Fe\,{\sc ii}]$1.644\,\mu$m emission. This indicates that the
relative contribution of the SNRs is nearly as large in the diffuse
background as it is in the radio SNR.
Interestingly, we find that most of the coincidences
with the radio SNRs occur in regions of  enhanced
[Fe\,{\sc ii}]$1.644\,\mu$m/Pa$\alpha$ line ratio, or conversely
we do not find a large number of coincidences in regions with
high Pa$\alpha$ flux (that is, H\,{\sc ii} regions).

\section{CALIBRATION OF THE SUPERNOVA RATE --
[Fe\,{\sc ii}] LUMINOSITY RELATION
}
The supernova rate in M82 has been estimated from number counts of the
radio supernova remnants and their ages to be  
$0.11\pm 0.05\,$yr$^{-1}$ (Huang et al. 1994). 
The total [Fe\,{\sc ii}]$1.644\,\mu$m luminosity in M82 is
$L$([Fe\,{\sc ii}]$)_{\rm total}=1.9 \times 10^{33}\,$W. Our value for the
[Fe\,{\sc ii}] luminosity of M82 is approximately 1.5 and 2  times higher
than the previous estimates by Greenhouse et al. (1997) and
Vanzi \& Rieke (1997), respectively. This difference is
most likely due to better extinction estimates made possible
by our NICMOS images. If the supernova rate in
M82 is $0.11\,$yr$^{-1}$, then we can derive the following calibration
for the supernova rate in terms of the [Fe\,{\sc ii}]$1.644\,\mu$m
luminosity:

\begin{equation}
{\rm (SNr)}_{\rm M82} = 0.6 \times \frac{L_{\rm [Fe\,{\sc ii}]}}{10^{34}
{\rm W}}\,\,\, ({\rm yr}^{-1})
\end{equation}

\noindent The total [Fe\,{\sc ii}]$1.644\,\mu$m luminosity (corrected for
extinction) measured in NGC~253 is $6.1\times 10^{32}\,{\rm W}$.
An upper limit on the supernova rate
from radio supernova counts is $<0.1-0.3\,{\rm yr}^{-1}$
(Ulvestad \& Antonucci 1997). A rate of $0.08\,{\rm yr}^{-1}$
has been calculated assuming that the non-thermal
radio flux is entirely due to synchrotron emission
(Van Buren \& Greenhouse 1994); this value is consistent
with the similar number of compact radio sources
in this galaxy compared with M82. The calibration of supernova rate
in terms of [Fe\,{\sc ii}]$1.644\,\mu$m luminosity for this galaxy is then

\begin{equation}
{\rm (SNr)}_{\rm NGC253} = 1.3 \times \frac{L_{\rm [Fe\,{\sc ii}]}}{10^{34}
{\rm W}}\,\,\, ({\rm yr}^{-1})
\end{equation}

\noindent Since M82 is the more thoroughly studied galaxy in the radio,
we give higher weight to the calibration for it in averaging
these results to estimate

\begin{equation}
{\rm (SNr)} = 0.8 \times \frac{L_{\rm [Fe\,{\sc ii}]}}{10^{34}
{\rm W}}\,\,\, ({\rm yr}^{-1})
\end{equation}

In the above calculations, we have included the entire
[Fe\,{\sc ii}] luminosity, since that is the parameter
usually measured in a distant galaxy. However, in these
nearby examples, we can compare the [Fe\,{\sc ii}] emission
produced in H\,{\sc ii} regions with that from supernovae.
We assume that the ratio for a typical H\,{\sc ii} region is
$f$([Fe\,{\sc ii}]$1.644\,\mu$m$)_{\rm HII} \simeq 6.6 \times 10^{-3}
\,f({\rm Pa}\alpha)$ (e.g., Orion). We obtained that the [Fe\,{\sc ii}]
luminosity produced in H\,{\sc ii} regions accounts for  
less than 8\% the total [Fe\,{\sc ii}] luminosity in M82.

\section{SUMMARY AND CONCLUSIONS}

We have presented {\it HST}/NICMOS observations of the starburst galaxies M82
and NGC~253.
The morphology of the
[Fe\,{\sc ii}]$1.644\,\mu$m line emission in both galaxies displays two
components: (1) a disk component with a number of
superimposed compact sources  and (2) a diffuse component in the form of
filaments above and below the disk of the galaxy.
In M82 we find a remarkable similarity between the extended component,
the fingers or chimneys  above
and below the disk, of the radio and [Fe\,{\sc ii}] emissions.

In M82, we have detected in both Pa$\alpha$ and [Fe\,{\sc ii}] a 
nuclear ring of star formation. 
This ring, with an approximate diameter of 400\,pc, 
has also been observed in other
near- and mid-infrared hydrogen recombination lines and in the
[Ne\,{\sc ii}]$12.8\,\mu$m line. The peak of the $2.2\,\mu$m 
continuum brightness of M82 lies very close to (within 1\arcsec) the dynamical 
center, and corresponds
to the center of the ring of star formation. In NGC~253, bright Pa$\alpha$
emission stems from the near-infrared continuum peak and extends in a
plume-like morphology along the disk of the galaxy. There is also a
diffuse Pa$\alpha$ component above and below the plane of galaxy, however, much
fainter than that in M82.

Prior to calibrating the surpernova rate in terms of the [Fe\,{\sc ii}]
emission, we have derived the extinction to the central regions of
NGC~253 and M82. To assess the extinction correction, 
we have compared  the 
number of ionizing photons infered from the 
extinction-corrected Pa$\alpha$ maps with 
other estimates based on longer wavelength indicators, and found
a good agreement.

Since the [Fe\,{\sc ii}] emission in starburst galaxies is 
thought to be associated with
supernova activity, we have cross-correlated the positions 
(with a spatial resolution of a few tenths of arcseconds) of the radio 
SNRs and the identified compact 
[Fe\,{\sc ii}] sources in M82 and NGC~253.
We have found that approximately $30-50$\% of the radio SNRs 
have an [Fe\,{\sc ii}] counterpart. However, we cannot
associate an [Fe\,{\sc ii}] source to each radio SNR or vice versa.
This is well understood in terms of
the different ages traced by the radio SNRs (a few hundred years) and
the [Fe\,{\sc ii}] stage of SNRs ($ \lesssim 10^4\,$yr).
The compact [Fe\,{\sc ii}] emitting sources in M82 and NGC~253
only contribute to some $20-30\%$  of the total [Fe\,{\sc ii}]$1.644\,\mu$m
emission.  However, much of the remaining emission in the disk
of the two galaxies is most likely produced by SNRs that expanded and
merged into a general ISM a few $10^4\,$yr ago. Photometry on the
[Fe\,{\sc ii}]$1.644\,\mu$m emission arising from the disk of M82 shows
that as much as $70\%$ of the total [Fe\,{\sc ii}]$1.644\,\mu$m emission
(corrected for extinction) is associated with SNRs. The remaining diffuse
[Fe\,{\sc ii}] emission perpendicular to the disk of M82 
seems to be related to  the superwind (via expulsion of material from 
the central starburst), as argued from the radio emission morphology
(Wills et al. 1999).

All these arguments support the supernova origin for the observed 
near-infrared [Fe\,{\sc ii}] line emission in starburst galaxies. 
Thus we have recalibrated the supernova rate in terms
of the [Fe\,{\sc ii}] luminosity
relation for starburst galaxies 
using our newly measured (corrected for extinction)
[Fe\,{\sc ii}]$1.644\,\mu$m luminosities of M82 and NGC~253.

\section*{Acknowledgments}

We would like to thank Drs. Karen Wills, T. Muxlow and A. Pedlar for kindly
providing us with the radio images for comparison with the NICMOS 
images, and the list of radio SNR. We are also grateful to Dr. Ernesto 
Oliva for enlightening discussions, and an anonymous referee for 
constructive comments that helped improve the paper.

During the course of this work AA-H was supported by the National
Aeronautics and Space Administration on grant NAG 5-3042 through the
University of Arizona. The work was also partially supported by the
National Science Foundation under grant AST-95-29190.

{}

\clearpage

\begin{figure}
\figurenum{1a}
%\plotfiddle{figure1a.ps}{425pt}{-90}{70}{70}{-260}{450}
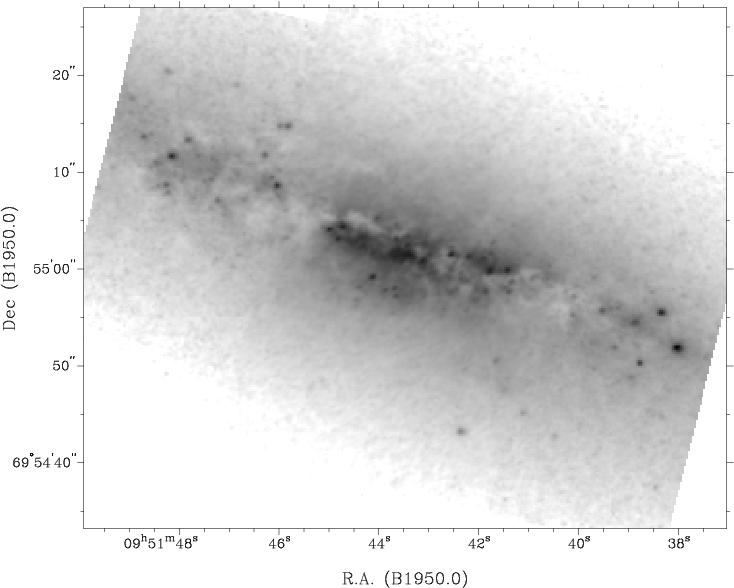
\vspace{2cm}
\caption{M82: continuum emission at $1.66\,\mu$m (NIC3 F166N image),
displayed on a logarithmic scale. }
\end{figure}

\begin{figure}
\figurenum{1b}
%\plotfiddle{figure1b.ps}{425pt}{-90}{70}{70}{-260}{450}
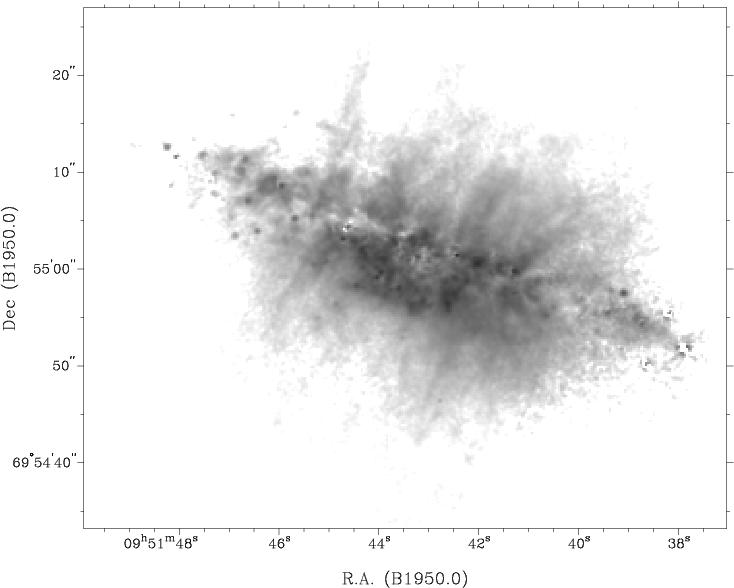
\vspace{2cm}
\caption{M82: [Fe\,{\sc ii}]$1.644\,\mu$m line
emission (continuum subtracted NIC3 F164N image), displayed
on a logarithmic scale.}
\end{figure}

\begin{figure}
\figurenum{2a}
\plotfiddle{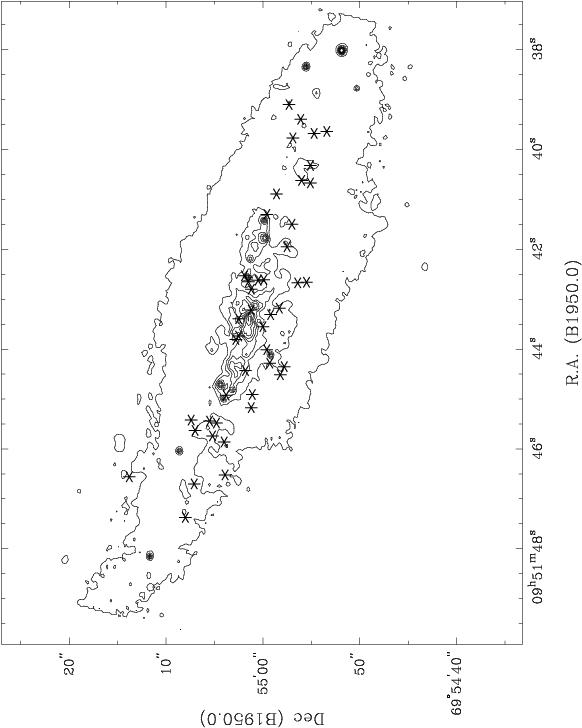}{425pt}{-90}{70}{70}{-260}{480}
\vspace{-3cm}
\caption{Contour map of the M82 continuum emission at $1.66\,\mu$m
on a linear scale. The asterisks are the positions of the radio SNRs 
detected in radio maps (Karen Wills, private communication
2000, see text for details).}
\end{figure}

\begin{figure}
\figurenum{2b}
\plotfiddle{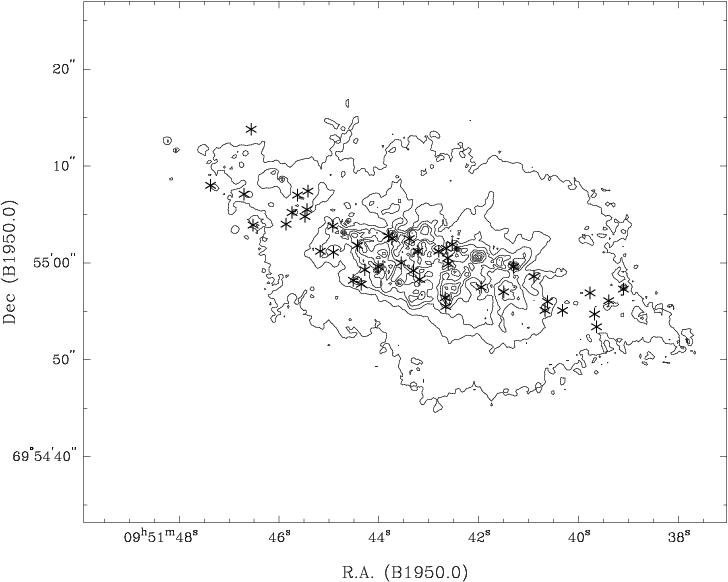}{425pt}{-90}{70}{70}{-260}{480}
\vspace{-3cm}
\caption{Contour map of the M82 [Fe\,{\sc ii}]$1.644\,\mu$m 
line emission
(continuum subtracted) on a linear scale. Symbols as in Figure~2a.}
\end{figure}

\begin{figure}
\figurenum{3a}
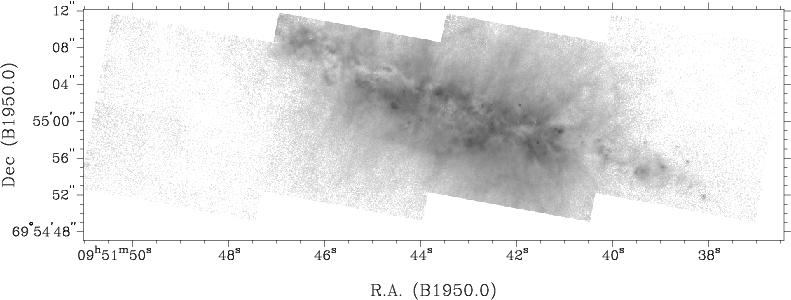
\vspace{2cm}
%\plotfiddle{figure3a.ps}{425pt}{-90}{80}{80}{-300}{400}
\caption{M82: continuum emission at $1.90\,\mu$m 
(NIC2 F190N images) displayed on a logarithmic scale.}
\end{figure}

\begin{figure}
\figurenum{3b}
figure3b.jpeg
\vspace{2cm}
%\plotfiddle{figure3b.ps}{425pt}{-90}{80}{80}{-300}{400}
\caption{M82: Pa$\alpha$ ($1.87\,\mu$m) line emission map (continuum
subtracted NIC2 F187N images) displayed on a logarithmic scale.}
\end{figure}

\begin{figure}
\figurenum{4a}
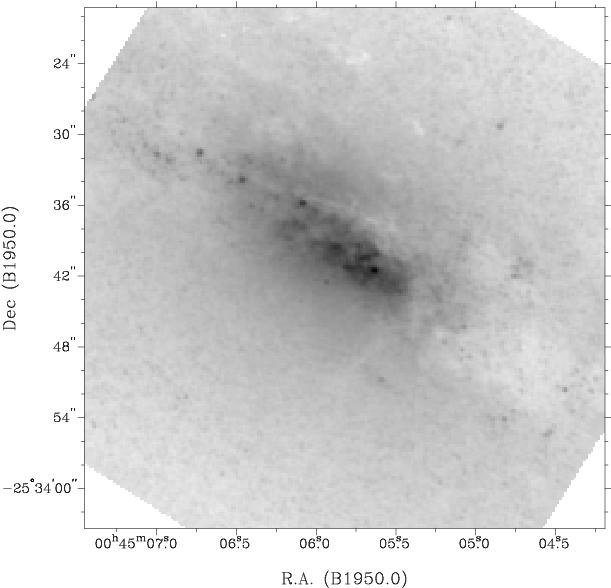
\vspace{2cm}
%\plotfiddle{figure4a.ps}{425pt}{-90}{70}{70}{-260}{450}
\caption{NGC~253: continuum emission 
at $1.66\,\mu$m (NIC3 F166N image), displayed on a logarithmic scale.}
\end{figure}

\begin{figure}
\figurenum{4b}
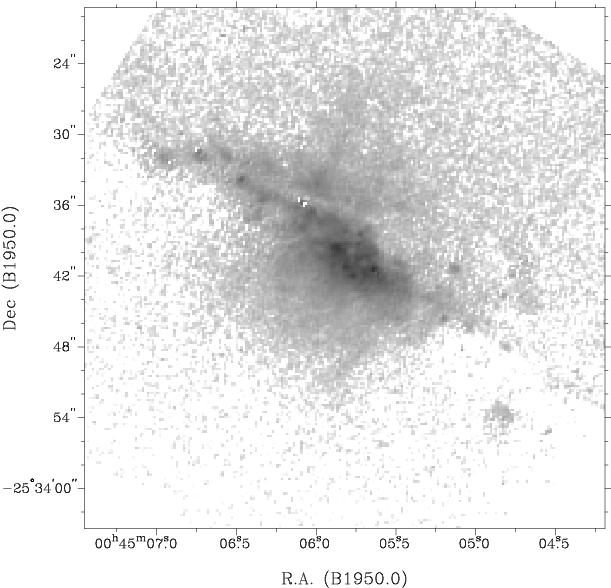
\vspace{2cm}
%\plotfiddle{figure4b.ps}{425pt}{-90}{70}{70}{-260}{450}
\caption{NGC~253: [Fe\,{\sc ii}]$1.644\,\mu$m line
emission (continuum subtracted NIC3 F164N image), displayed
on a logarithmic scale.}
\end{figure}

\begin{figure}
\figurenum{5a}
\plotfiddle{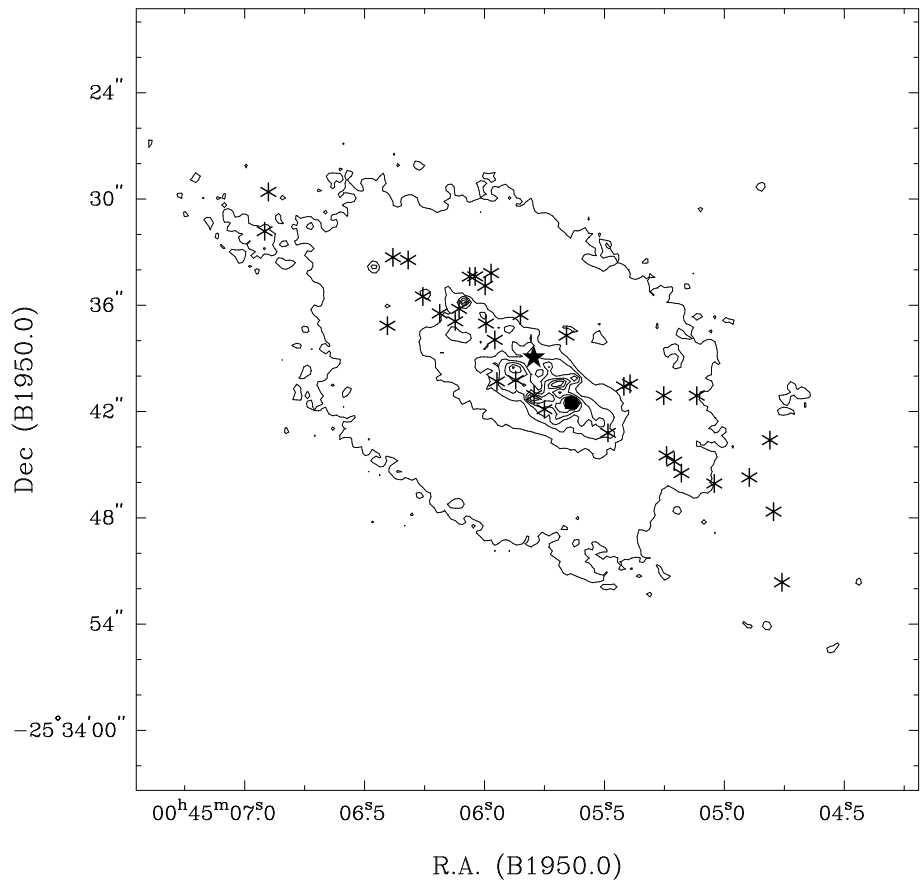}{425pt}{-90}{70}{70}{-260}{480}
\vspace{-3cm}
\caption{Contour map of the NGC~253 continuum 
emission at $1.66\,\mu$m on a linear scale. The asterisks 
are the positions of the 6\,cm radio SNRs from Ulvestad \&
Antonucci (1991) with $S_{\rm peak} > 1\,$mJy, excluding
those sources classified as H\,{\sc ii}
regions in Ulvestad \& Antonucci (1997). The filled 
triangle is the position of the peak of the near-infrared emission, 
whereas the filled star symbol is the position of the non-thermal
radio source (see text for details).}
\end{figure}

\begin{figure}
\figurenum{5b}
\plotfiddle{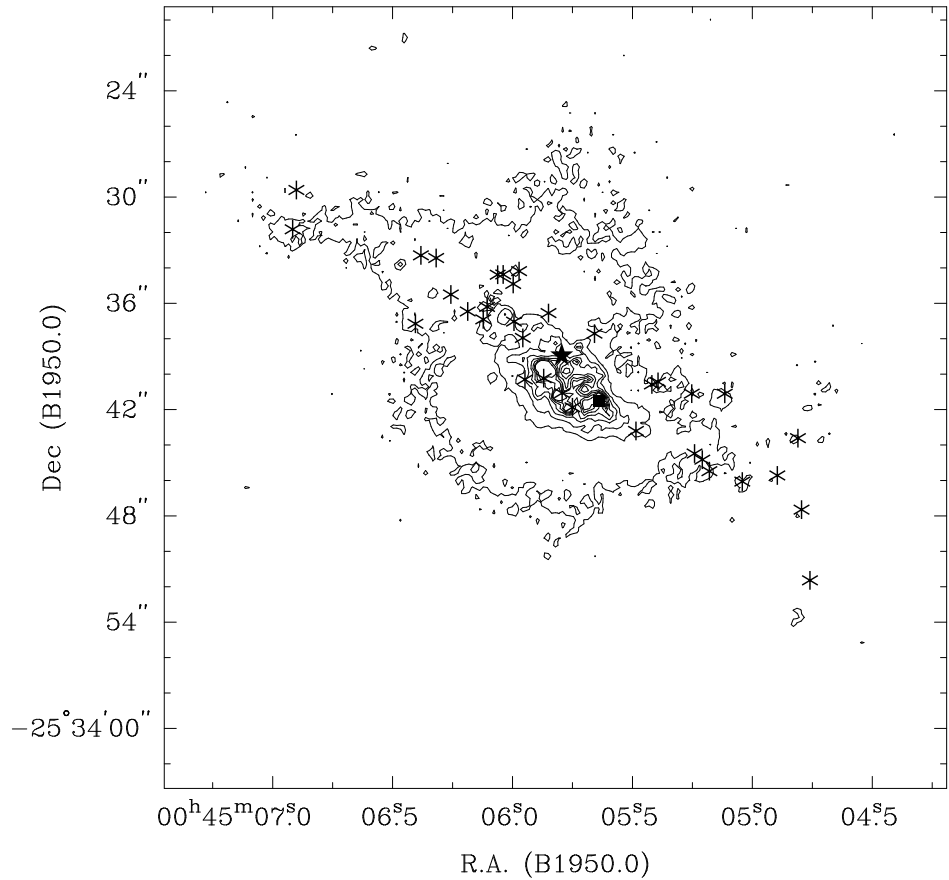}{425pt}{-90}{70}{70}{-260}{480}
\vspace{-3cm}
\caption{Contour map of the NGC~253 
[Fe\,{\sc ii}]$1.644\,\mu$m line emission
(continuum subtracted) on a linear scale. Symbols as in Figure~5a.}
\end{figure}

\begin{figure}
\figurenum{6a}
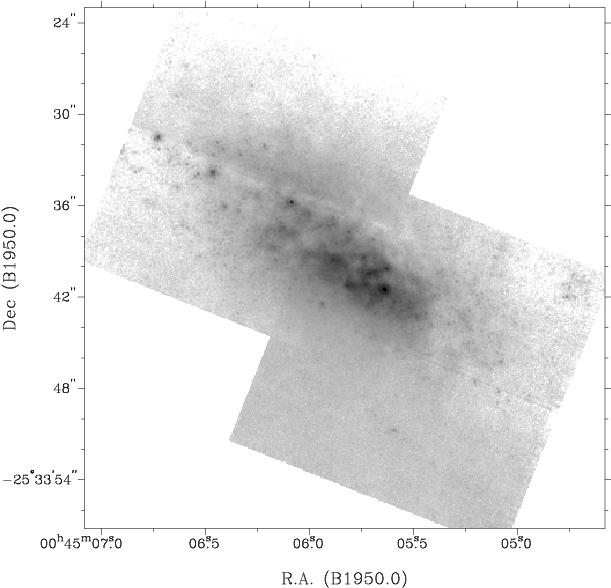
\vspace{2cm}
%\plotfiddle{figure6a.ps}{425pt}{-90}{80}{80}{-300}{450}
\caption{NGC~253: continuum emission 
at $1.90\,\mu$m (NIC2 F190N images) displayed on a logarithmic scale.}
\end{figure}

\begin{figure}
\figurenum{6b}
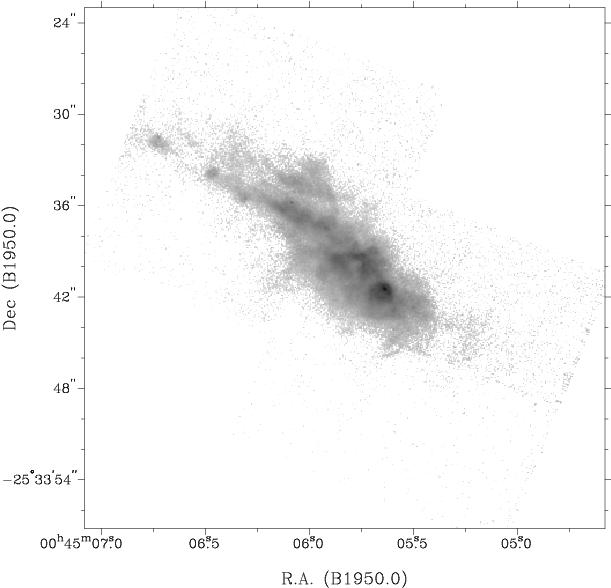
\vspace{2cm}
%\plotfiddle{figure6b.ps}{425pt}{-90}{80}{80}{-300}{450}
\caption{NGC~253: Pa$\alpha$ ($1.87\,\mu$m) 
line emission map (continuum
subtracted NIC2 F187N images) displayed on a logarithmic scale.}
\end{figure}

\begin{figure}
\figurenum{7a}
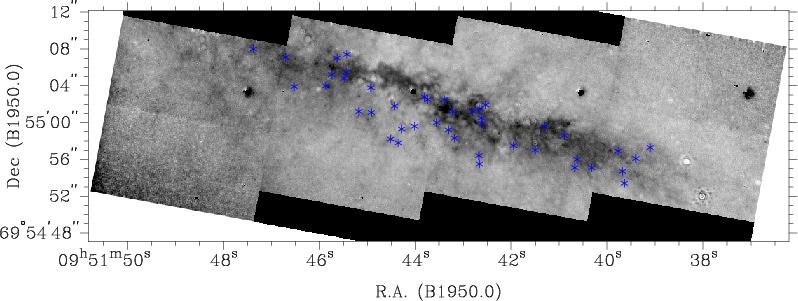
\vspace{2cm}
%\plotfiddle{figure7a.ps}{425pt}{-90}{80}{80}{-300}{400}
\caption{M82: $H-K$ color map (NIC2 F160W - NIC2 F222M) 
which represents
the extinction to the stars. Symbols are as in Figure~2a. Dark regions 
indicate regions of higher
extinction. The circular black spot on the upper right corners of the
individual images is the coronographic hole.}
\end{figure}

\begin{figure}
\figurenum{7b}
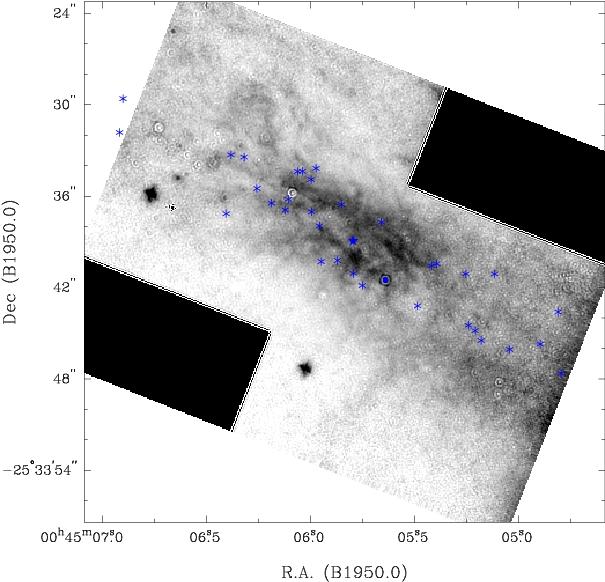
\vspace{2cm}
%\plotfiddle{figure7b.ps}{425pt}{-90}{80}{80}{-300}{450}
\caption{NGC~253: $H-K$ color map (NIC2 F160W - NIC2 F222M).
Symbols are as in Figure~5a. The black spot on left bottom
corners of the individual images is the coronographic hole. }

\end{figure}

%\newpage

\begin{figure}
\figurenum{8a}
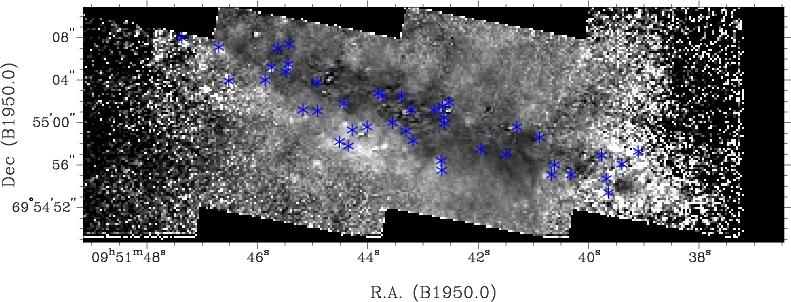
\vspace{2cm}
%\plotfiddle{figure8a.ps}{425pt}{-90}{80}{80}{-300}{400}
\caption{
M82: [Fe\,{\sc ii}]$1.644\,\mu$m/Pa$\alpha$ line ratio
map. Lighter colors indicate regions of enhanced
[Fe\,{\sc ii}]$1.644\,\mu$m/Pa$\alpha$ line ratios. Superimposed
are the positions of the radio SNRs as in Figure~2a.}
\end{figure}

\begin{figure}
\figurenum{8b}
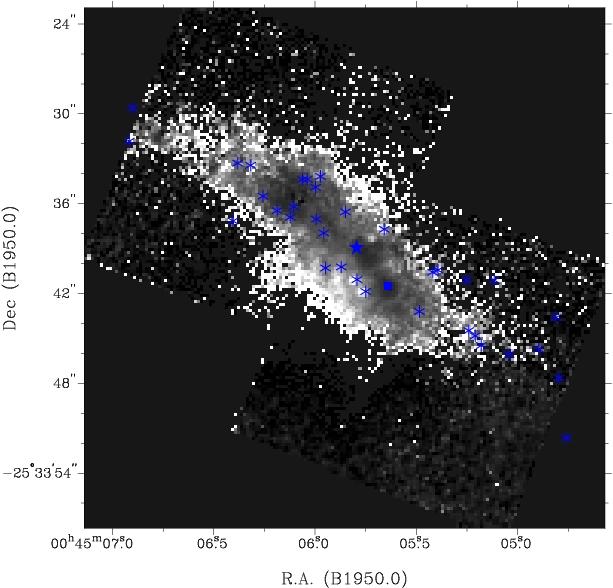
\vspace{2cm}
%\plotfiddle{figure8b.ps}{425pt}{-90}{80}{80}{-300}{450}
\caption{NGC~253 [Fe\,{\sc ii}]$1.644\,\mu$m/Pa$\alpha$ line ratio
map. Colors as in Figure~8a. Superimposed  are the positions
of the radio SNRs as in Figure~5a.}
\end{figure}

%\newpage

\begin{figure}
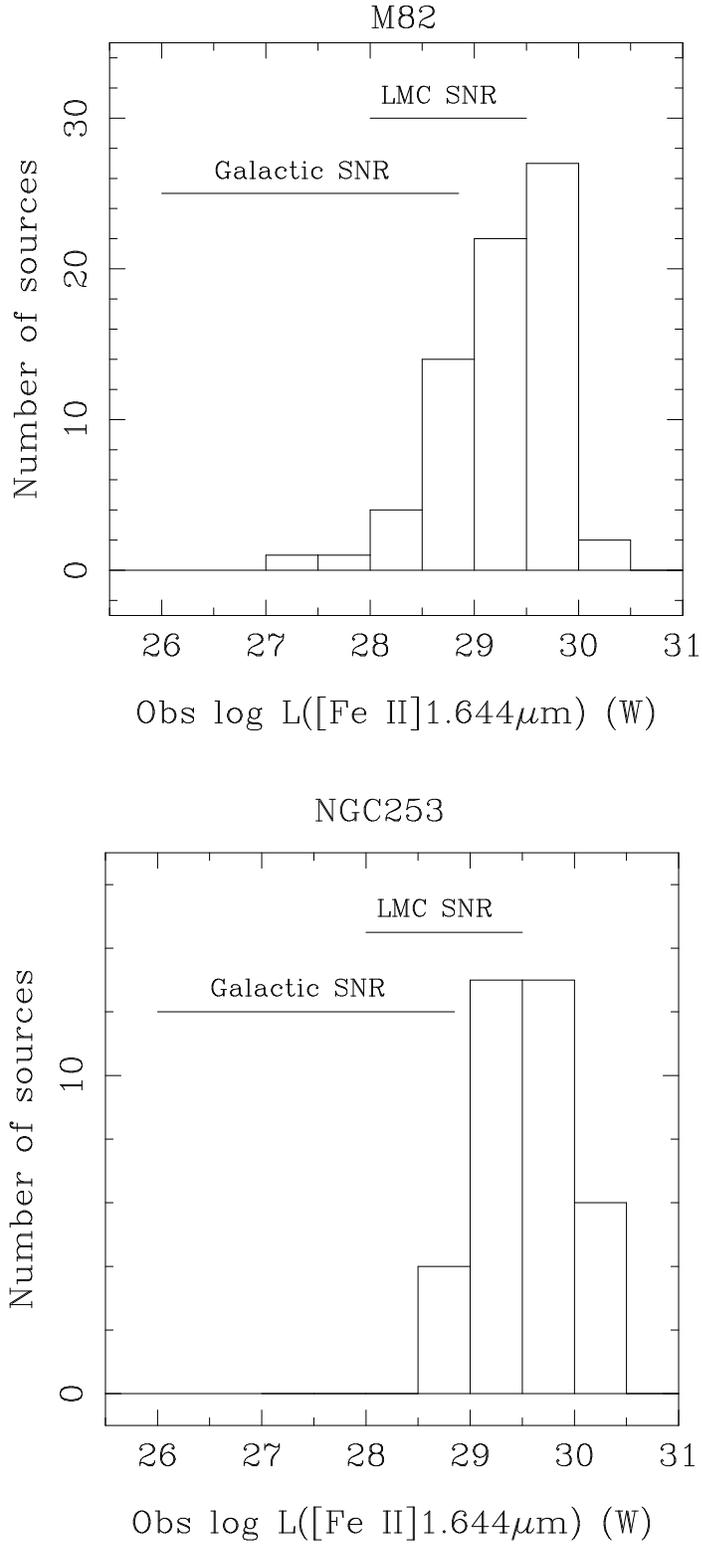

\figurenum{9}
\plotfiddle{figure9a.ps}{425pt}{0}{100}{100}{-210}{120}
\plotfiddle{figure9b.ps}{425pt}{0}{100}{100}{-210}{250}
\vspace{-10cm}
\caption{Distributions of the
[Fe\,{\sc ii}]$1.644\,\mu$m luminosities (not corrected
for extinction) of the compact sources in M82 and NGC~253
measured through a $0.8\arcsec$ diameter aperture.}
\end{figure}

%\clearpage

\begin{figure}
\figurenum{10}
\plotfiddle{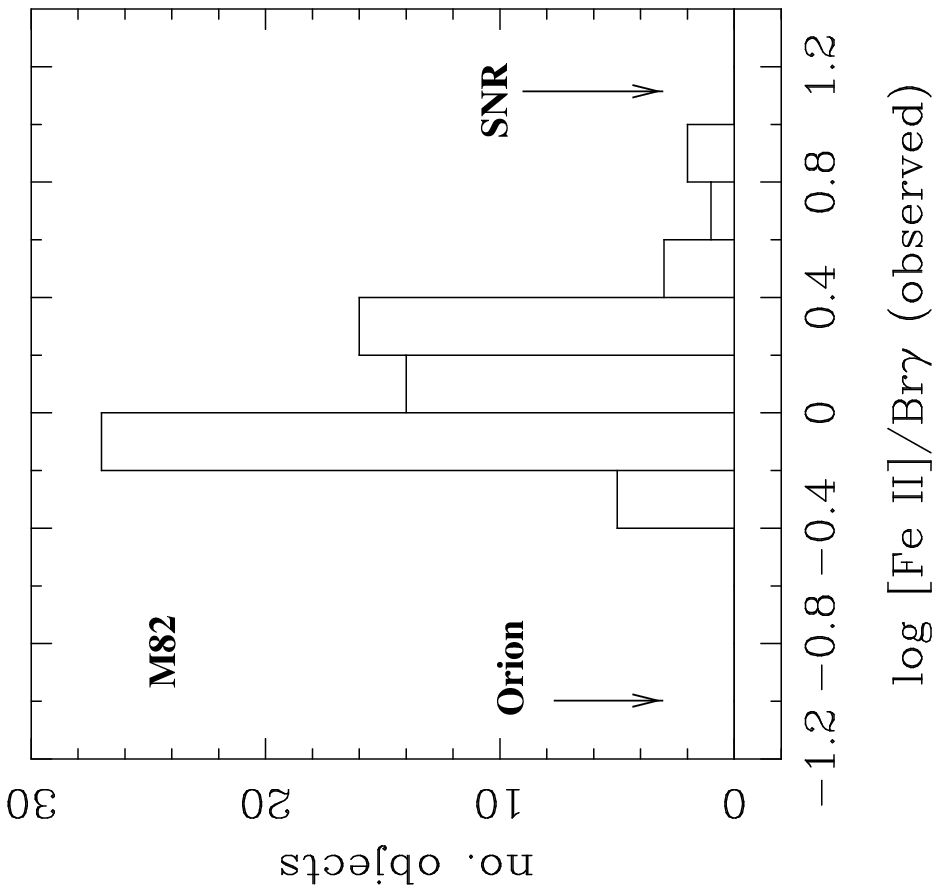}{425pt}{-90}{100}{100}{-190}{440}
\plotfiddle{figure10b.ps}{425pt}{0}{100}{100}{-210}{270}
\vspace{-11cm}
\caption{Histograms of the observed
[Fe\,{\sc ii}]$1.644\,\mu$m/Br$\gamma$  line ratios of the
compact [Fe\,{\sc ii}] sources.}
\end{figure}

\end{document}